\documentclass[a4paper,10pt,twoside]{cpc-hepnp}
\usepackage{multicol}
\usepackage{graphicx}
\usepackage{booktabs}
\usepackage{amssymb,bm,mathrsfs,bbm,amscd}
\usepackage[tbtags]{amsmath}
\usepackage{lastpage}
\usepackage{CJK}

\begin{document}

\fancyhead[c]{\small To appear in ¡®Chinese Physics C¡¯} \fancyfoot[C]{\small -\thepage}

\title{NLO QCD Corrections to $\eta_c+h_c(1P)/\psi_{1,2}(1D)$ Production at B-factories\thanks{This work was supported in part by the Ministry of Science and Technology of the People's Republic of China (2015CB856703), and by the National Natural Science Foundation of China(NSFC) under the grants 11175249 and 11375200.}}

\author{%
Long-Bin Chen$^{1,2; 1)}$\email{chenglogbin10@mails.ucas.ac.cn}%
\quad Jun Jiang$^{1; 2)}$\email{jiangjun13b@mails.ucas.ac.cn}%
\quad Cong-Feng Qiao$^{1,2; 3)}$\email{qiaocf@ucas.ac.cn, corresponding author}%
}

\maketitle

\address{%
$^1$School of Physics, University of Chinese Academy of Sciences, YuQuan Road 19A, Beijing 100049, China\\
$^2$CAS Center for Excellence in Particle Physics, Beijing 100049, China\\
}

\begin{abstract}
We calculate the next-to-leading order (NLO) quantum chromodynamics (QCD) corrections to double charmonium production processes $e^+e^-\to\gamma^*\to\eta_c+h_c(1P)/\psi_{1,2}(1D)$ within the non-relativistic QCD (NRQCD) factorization framework.
We find that the corrections to $\eta_c+h_c(1P)$ production are positive, while those to $\eta_c+\psi_{1,2}(1D)$ are negative. Unlike the $J/\psi + \eta_c$ case, all the corrections here are not large. Uncertainties in the renormalization scale, quark mass and running energy of center-of-mass are discussed, and the scale dependence of these processes is found to be greatly reduced with the NLO QCD corrections.
\end{abstract}



\begin{multicols}{2}

\section{Introduction}

Once, the double charmonium inclusive(exclusive) processes $e^+e^-\to J/\psi+c\bar{c}(\eta_c)$ encouraged many investigations because of large discrepancies between leading order (LO) calculations \cite{inclusive, exclusive1, exclusive2, exclusive3} and experimental results at B-factories \cite{Abe:2002rb, Aubert:2005tj}. It was found that this problem may
be somehow remedied by next-to-leading order (NLO) Quantum Chromodynamics (QCD) corrections \cite{Zhang:2005cha, Zhang:2006ay, Gong:2007db, Gong:2009ng, Dong:2012xx} in the framework of non-relativistic QCD (NRQCD) \cite{Bodwin:1994jh}, which indicates that the NLO corrections might be significant to LO results. In order to further understand the NLO properties and investigate the convergence of perturbative expansion for heavy quarkonium production and decays in the NRQCD formalism, we need to study more heavy quarkonium production and decay processes.

There have been great advances in recent years in the calculation of radiative corrections to charmonium inclusive and exclusive production and decays. The LO estimations for $h_c$ inclusive production at B-factories were given in Refs. \cite{Jia:2012qx, Wang:2012tz}. Recently, a complete NLO calculation for the P-wave charmonium $\chi_{cJ}(^3P^{[1]}_J, ^3S^{[8]}_1)$ inclusive production processes $e^+ e^- \to \chi_{cJ} + c\bar{c}/gg/q\bar{q} (q=u,d,s)$ was carried out \cite{Chen:2014ahh}.
Though so far there are insufficient data on the double excited charmonium exclusive processes $e^+e^-\to \gamma^*\to H_1+H_2$\footnote{To guarantee C-parity conservation, charmonium $H_1$ should have different C-parity from $H_2$ since $C(\gamma^*)=-$. However in the case of $H_1+H_2$ production through two photons, the two charmoniums should have the same C-parity \cite{exclusive3}.}, theoretical studies have already started \cite{exclusive2, excited}, and even the NLO corrections for $J/\psi+\chi_{cJ}$ production at B-factories \cite{Dong:2011fb, Wang:2011qg} have been performed.

In this work, we calculate the NLO QCD corrections for
$e^+e^-\rightarrow\eta_c+h_c(1P)/\psi_{1,2}(1D)$ processes within the NRQCD formalism. The $h_c(1^1P_1, J^{PC}=1^{+-})$ state was first observed by experiment E760 at Fermilab \cite{Armstrong:1992ae}, and its C-parity was established by radiative decay $h_c\to\eta_c\gamma$ \cite{Ablikim:2010rc}. The $\psi(3770)$ has been identified as $\psi_1(1^3D_1, J^{PC}=1^{--})$ since its parameters are consistent with the expectaions \cite{psi1}. Note, the $\psi(3836)$ was once considered to be the quark model state $\psi_2(1^3D_2, J^{PC}=2^{--})$ \cite{Antoniazzi:1993jz}, however, the $\psi(3823)$ recently observed by Belle \cite{Bhardwaj:2013rmw} and BESIII \cite{Ablikim:2015dlj} turns to be the $\psi_2(1^3D_2)$ state. For simplicity, we denote $\psi_{1,2}(1^3D_{1,2})$ states as $\psi_{1,2}$ throughout this paper. After the running of super B-factories in future, the processes considered in this study might be observed, and a precise evaluation is hence necessary.

The rest of this paper is organized as follows: Section 2
presents our formalism and calculation method, numerical
evaluation and some discussion of the results are given in Section 3, and
Section 4 gives a summary and conclusions.

\section{Calculation Scheme Description and Formalism}

In our calculation the Mathematica package FeynArts \cite{feynarts} was applied to generate all the Feynman diagrams and amplitudes of partons. The standard projection operators for charmonia may be expressed as \cite{bodwpe}:
\begin{eqnarray}
\Pi_{0,1}&=&\frac{1}{4\sqrt{2}E(E+m_c)} (\not\!\bar{p}-m_c)\not\!\epsilon_{0,1} (\not\!P+2E)(\not\!p+m_c)\nonumber\\ &&\otimes\frac{\bf{1}}{\sqrt{N_c}}\ .
\end{eqnarray}
Here, $\Pi_{0}$ corresponds to the spin-singlet charmonia with $\not\!\!\epsilon_{0}=\gamma_5$,
while $\Pi_{1}$ corresponds to spin-triplet states with $\not\!\!\epsilon_{1}=\not\!\!\epsilon^*$, the spin polarization vector.
$p=\frac{P}{2}+q$ and $\bar{p}=\frac{P}{2}-q$ are the momenta of the quark and antiquark within the charmonium, respectively.
$P$ denotes the momentum of quarkonium and $q$ is the relative momentum between quarks inside the quarkonium. ${\bf{1}}$ stands for the unit matrix in color space.
\begin{center}
\includegraphics[width=7.5cm]{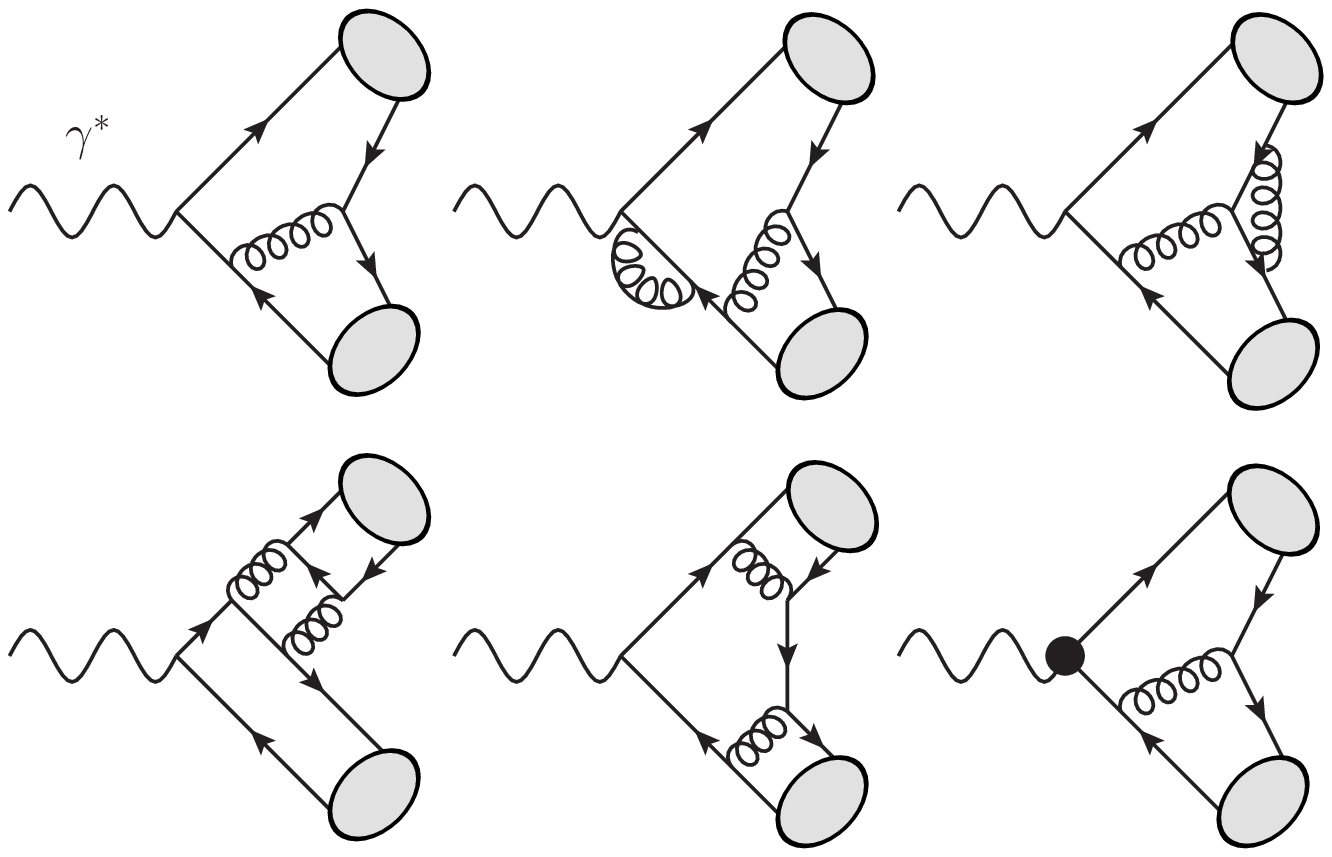}
\figcaption{\label{fig1} The typical Feynman diagrams of processes $\gamma^*\rightarrow\eta_c+h_c(1P)/\psi_{1,2}(1D)$ in NLO QCD.}
\end{center}

Using the method described in Ref. \cite{exclusive2}, and applying
FeynCalc \cite{Feyncalc} to assist the calculation of amplitudes,
one can readily obtain the tree-level results. For one-loop
QCD corrections, the representative Feynman diagrams are shown in Fig. 1. We first used FeynCalc to trace the Dirac matrix chains as well as the color matrices, and to perform the derivative on momentum $q$. Next, the package \$Apart \cite{apart} was employed to reduce the propagators of each one-loop diagram into linearly independent ones. Then, we applied the package FIRE \cite{fire} to reduce all the one-loop integrals into several master-integrals($A_0, B_0, C_0, D_0$). Finally, package LoopTools \cite{looptools} was employed to evaluate the scalar master-integrals numerically.

Throughout our calculation, we adopted Feynman gauge, and took the conventional dimensional regularization with $D=4-2\epsilon$ to regularize the divergences. The ultraviolet divergences are canceled by the counter terms and the infrared divergences in the short distance coefficients cancel each other, which confirms the NRQCD factorization for
$e^+e^-\rightarrow \eta_c+h_c(1P)/\psi_{1,2}(1D)$ processes at NLO level.
When handling the counter terms, we found that terms related to $Z_3$, the renormalization constant corresponding to the gluon field, vanish in the end. The renormalization constant $Z_g$, corresponding to the strong coupling constant $\alpha_s$, was computed in the modified-minimal-subtraction $(\overline{MS})$ scheme, while $Z_2$ and $Z_m$, corresponding respectively to the quark field and quark mass, were in the on-shell $(OS)$ scheme. Eventually, the expressions for the relevant renormalization constants read:
\begin{eqnarray}
  \delta Z_g^{\overline{\rm MS}}&=&-\frac{\beta_0}{2}\,
  \frac{\alpha_s}{4\pi}
  \left[\frac{1}{\epsilon_{\rm UV}} -\gamma_E + \ln(4\pi)
  \right],\nonumber\\
  \delta Z_2^{\rm OS}&=&-C_F\frac{\alpha_s}{4\pi}
  \left[\frac{1}{\epsilon_{\rm UV}}+\frac{2}{\epsilon_{\rm IR}}
  -3\gamma_E+3\ln\frac{4\pi\mu^2}{m^2}+4\right],\nonumber\\
  \delta Z_m^{\rm OS}&=&-3C_F\frac{\alpha_s}{4\pi}
  \left[\frac{1}{\epsilon_{\rm
  UV}}-\gamma_E+\ln\frac{4\pi\mu^2}{m^2} +\frac{4}{3}\right].
\end{eqnarray}

\section{Numerical results and analysis}

Before carrying out numerical calculation, the input parameters
need to be fixed. The NLO running coupling constant
\begin{eqnarray}
 \frac{\alpha_s(\mu)}{4\pi}=\frac{1}{\beta_0 L}-\frac{\beta_1\ln
L}{\beta_0^3L^2}
\end{eqnarray}
was employed with $L=\ln(\mu^2/\Lambda_{QCD}^2)$,
$\beta_0=(11/3)C_A-(4/3)T_Fn_f$ and
$\beta_1=(34/3)C_A^2-4C_FT_Fn_f-(20/3)C_AT_Fn_f$. In numerical evaluation,
the effective quark flavor number $n_f=4$ was adopted and
$\Lambda_{QCD} = 0.297\textrm{GeV}$ \cite{pdg}. At leading order in relativistic expansion $M_{\eta_c/h_c/\psi_{1,2}} = 2m_c$, and hence the charm quark mass $m_c = 1.7 \pm 0.2$GeV was taken. The values of NRQCD matrix elements were evaluated from the Bunchm\"{u}ller-Tye potential model \cite{potential}, i.e.,
\begin{eqnarray}
\langle\mathcal{O}_1\rangle_{\eta_{c}} \approx \frac{N_c}{2\pi}|R_S(0)|^2&=&0.387\ \mathrm{GeV^3},\nonumber\\
\langle\mathcal{O}_1\rangle_{h_{c}} \approx \frac{3N_c}{2\pi}|R^{'}_P(0)|^2&=&0.107\ \mathrm{GeV^5},\nonumber\\
\langle\mathcal{O}_1\rangle_{\psi_{1,2}}\approx \frac{15N_c}{4\pi}|R^{''}_D(0)|^2&=&0.054\ \mathrm{GeV^7}.
\end{eqnarray}
\tabcaption{\label{tab:1} Cross sections $\sigma(\mu)$(fb) of
$e^+e^-\rightarrow \eta_c+h_c(1P)$ at leading
order and next-to-leading order at $\mu=2 m_c$ and $\sqrt{s}/2$. The errors are induced by
$m_c=1.7\pm 0.2$ GeV.}
\begin{center}
\renewcommand\arraystretch{2}
$e^+e^-\rightarrow \eta_c+h_c(1P)$
  \begin{tabular}{c|cc|c}
  \toprule $\ \ \sigma(\mu)$(fb)& LO & NLO & K-factor\\
    \hline
    $\mu = 2m_c$& $0.278^{-0.143}_{+0.315}$ & $0.331^{-0.154}_{+0.309}$ & $1.19^{+0.12}_{-0.11}$ \\
    $\mu = \sqrt{s}/2$& $0.207^{-0.107}_{+0.235}$ & $0.291^{-0.139}_{+0.287}$ & $1.40^{+0.11}_{-0.10}$ \\
  \bottomrule
  \end{tabular}
\end{center}
\tabcaption{\label{tab:2} Cross sections $\sigma(\mu)$(fb) of
$e^+e^-\rightarrow \eta_c+\psi_1(1D)$ at leading
order and next-to-leading order at $\mu=2 m_c$ and $\sqrt{s}/2$. The errors are induced by
$m_c=1.7\pm 0.2$ GeV.}
\begin{center}
\renewcommand\arraystretch{2}
$e^+e^-\rightarrow \eta_c+\psi_1(1D)$
  \begin{tabular}{c|cc|c}
  \toprule $\ \ \sigma(\mu)$(fb)& LO & NLO & K-factor\\
    \hline
    $\mu = 2m_c$& $0.217^{-0.124}_{+0.273}$ & $0.129^{-0.078}_{+0.181}$ & $0.596^{-0.046}_{+0.038}$ \\
    $\mu = \sqrt{s}/2$& $0.162^{-0.093}_{+0.203}$ & $0.144^{-0.085}_{+0.193}$ & $0.890^{-0.039}_{+0.033}$ \\
  \bottomrule
  \end{tabular}
\end{center}
\tabcaption{\label{tab:3} Cross sections $\sigma(\mu)$(fb) of
$e^+e^-\rightarrow \eta_c+\psi_2(1D)$ at leading
order and next-to-leading order at $\mu=2 m_c$ and $\sqrt{s}/2$. The errors are induced by
$m_c=1.7\pm 0.2$ GeV.}
\begin{center}
\renewcommand\arraystretch{2}
$e^+e^-\rightarrow \eta_c+\psi_2(1D)$
  \begin{tabular}{c|cc|c}
  \toprule $\ \ \sigma(\mu)$(fb)& LO & NLO & K-factor\\
    \hline
    $\mu = 2m_c$& $0.869^{-0.520}_{+1.247}$ & $0.605^{-0.341}_{+0.678}$ & $0.696^{+0.058}_{-0.090}$ \\
    $\mu = \sqrt{s}/2$& $0.648^{-0.388}_{+0.929}$ & $0.632^{-0.365}_{+0.785}$ & $0.976^{+0.050}_{-0.077}$ \\
  \bottomrule
  \end{tabular}
\end{center}

In our calculation, two typical renormalization scales were considered, therefore the corresponding values of running coupling constant are $\alpha_s(\mu=2m_c) = 0.235$ and $\alpha_s(\mu=\sqrt{s}/2) = 0.203$. The cross sections of $e^+e^-\rightarrow \eta_c+h_c(1P)/\psi_{1,2}(1D)$ are presented in Tabs. \ref{tab:1}-\ref{tab:3}, in which the errors are induced by varying $m_c$, and the K-factor is defined as
$\frac{\sigma_{NLO}} {\sigma_{LO}}$. These results indicate that:
\begin{description}
  \item[(1)] For $\eta_c+h_c(1P)$ production, the NLO correction enhances tree-level result, and the K-factor is bigger at large scale $\mu=\sqrt{s}/2$;
  \item[(2)] For $\eta_c+\psi_{1,2}(1D)$ productions, the NLO corrections are negative, and the K-factors are also bigger at large scale $\mu=\sqrt{s}/2$;
  \item[(3)] At LO, cross sections of these three processes decrease with the scale $\mu$ increasing from $2m_c$ to $\sqrt{s}/2$. The NLO result of $\eta_c+h_c(1P)$ also decrease with the scale, while the NLO results of $\eta_c+\psi_{1,2}(1D)$ have inverse relation with the energy scale in the discussed region;
  \item[(4)] The hierarchy $\sigma_{\psi_1}<\sigma_{h_c}<\sigma_{\psi_2}$ remains for both LO and NLO results at scales $\mu=2m_c$ and $\mu=\sqrt{s}/2$.
\end{description}

To illustrate the results in Tables \ref{tab:1}-\ref{tab:3} more clearly, we show the LO and NLO cross sections of $\eta_c+h_c(1P)/\psi_{1,2}(1D)$ production with different $m_c$ versus renormalization scale $\mu$ in Figs. \ref{fig2}-\ref{fig4} respectively. One may notice that the scale dependence of the NLO results is obviously depressed, and the convergence of pertubative expansion for these processes works well, whereas the cross sections are quite sensitive to the quark mass.
\begin{center}
\includegraphics[width=8.5cm]{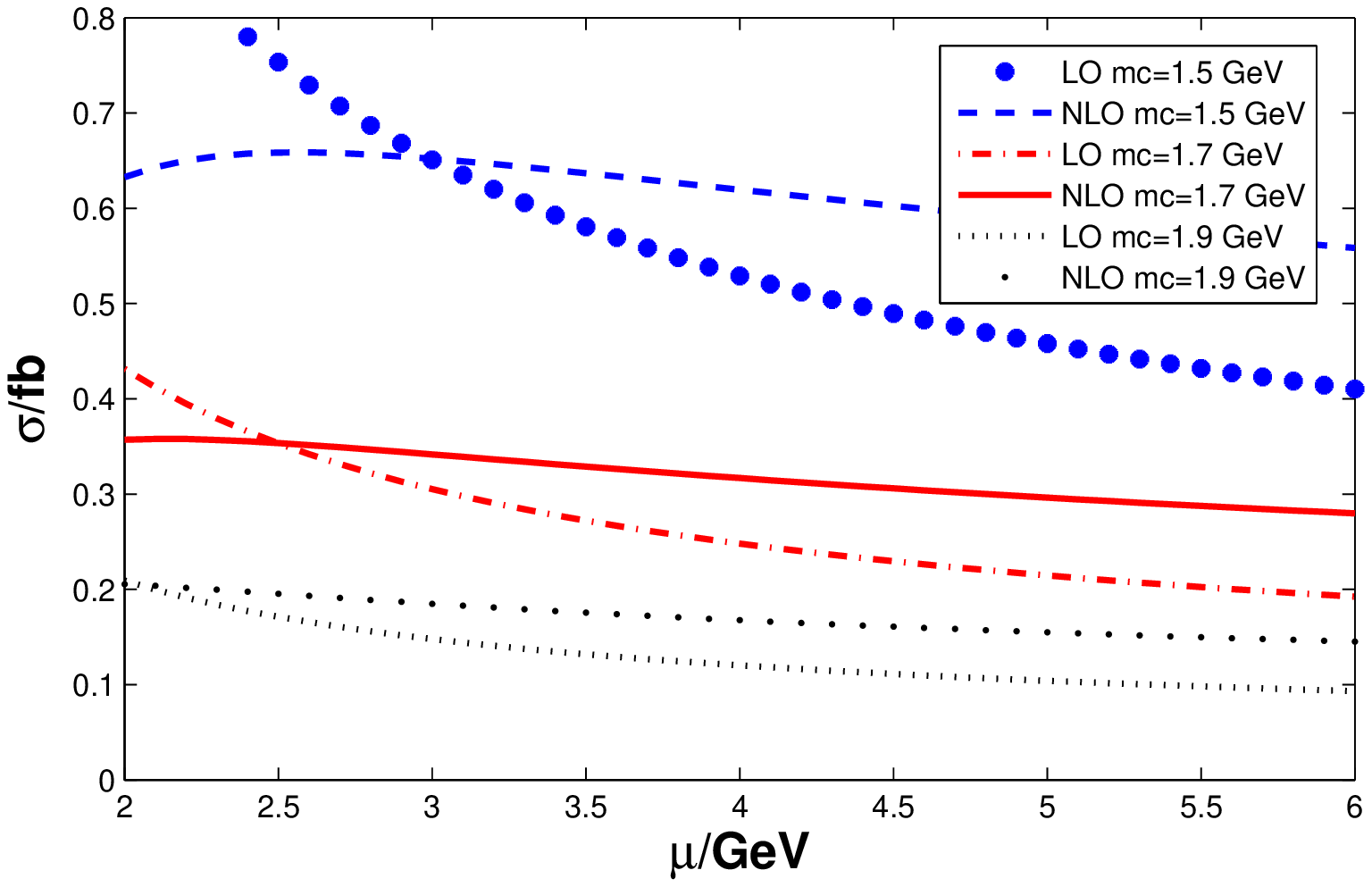}
\figcaption{\label{fig2} LO and NLO cross sections with different $m_c$ versus running scale $\mu$ for $e^+e^-\rightarrow\eta_c+h_c(1P)$.}
\end{center}
\begin{center}
\includegraphics[width=8.5cm]{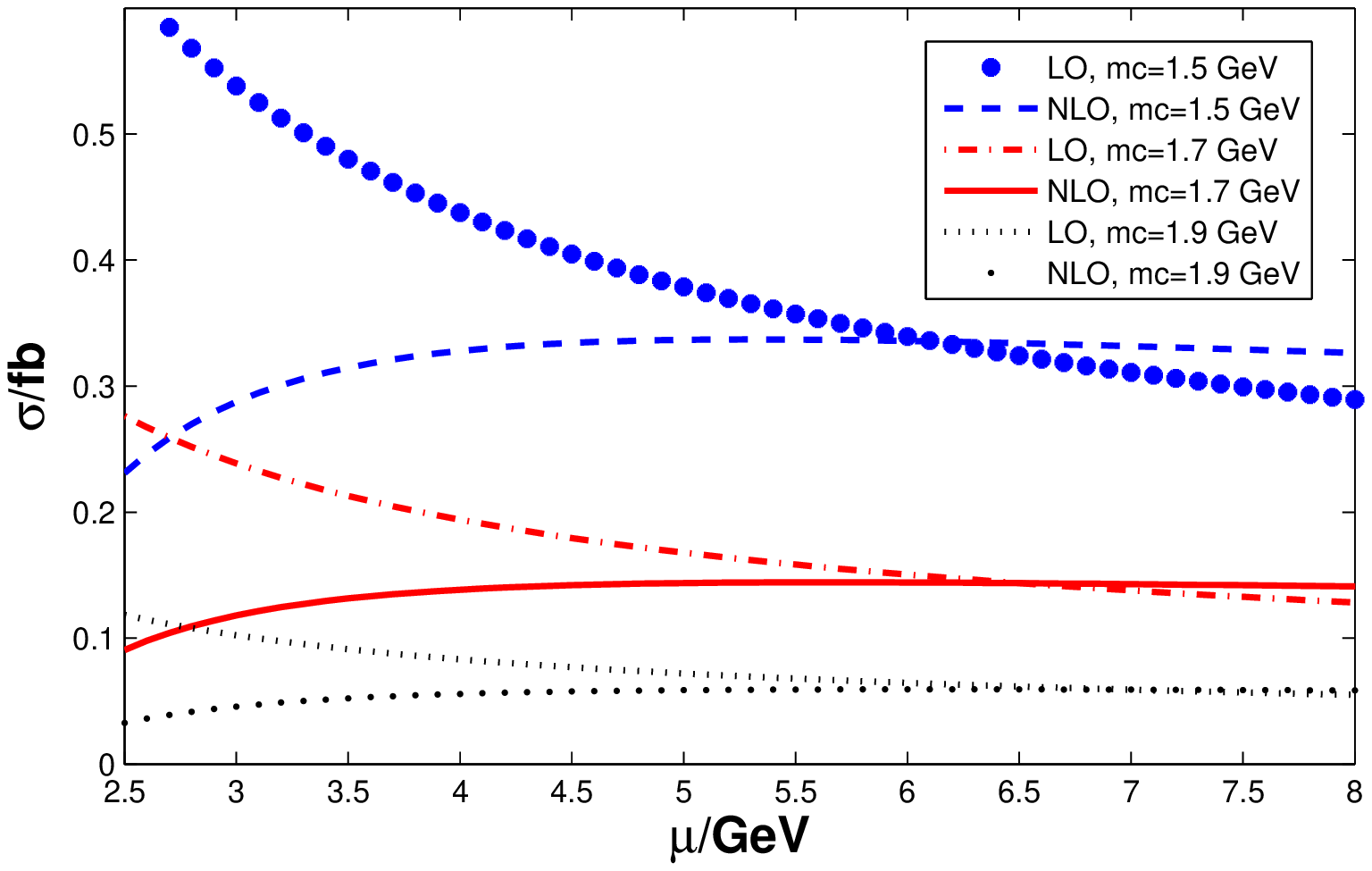}
\figcaption{\label{fig3} LO and NLO cross sections with different $m_c$ versus running scale $\mu$ for $e^+e^-\rightarrow\eta_c+\psi_{1}(1D)$.}
\end{center}
\begin{center}
\includegraphics[width=8.5cm]{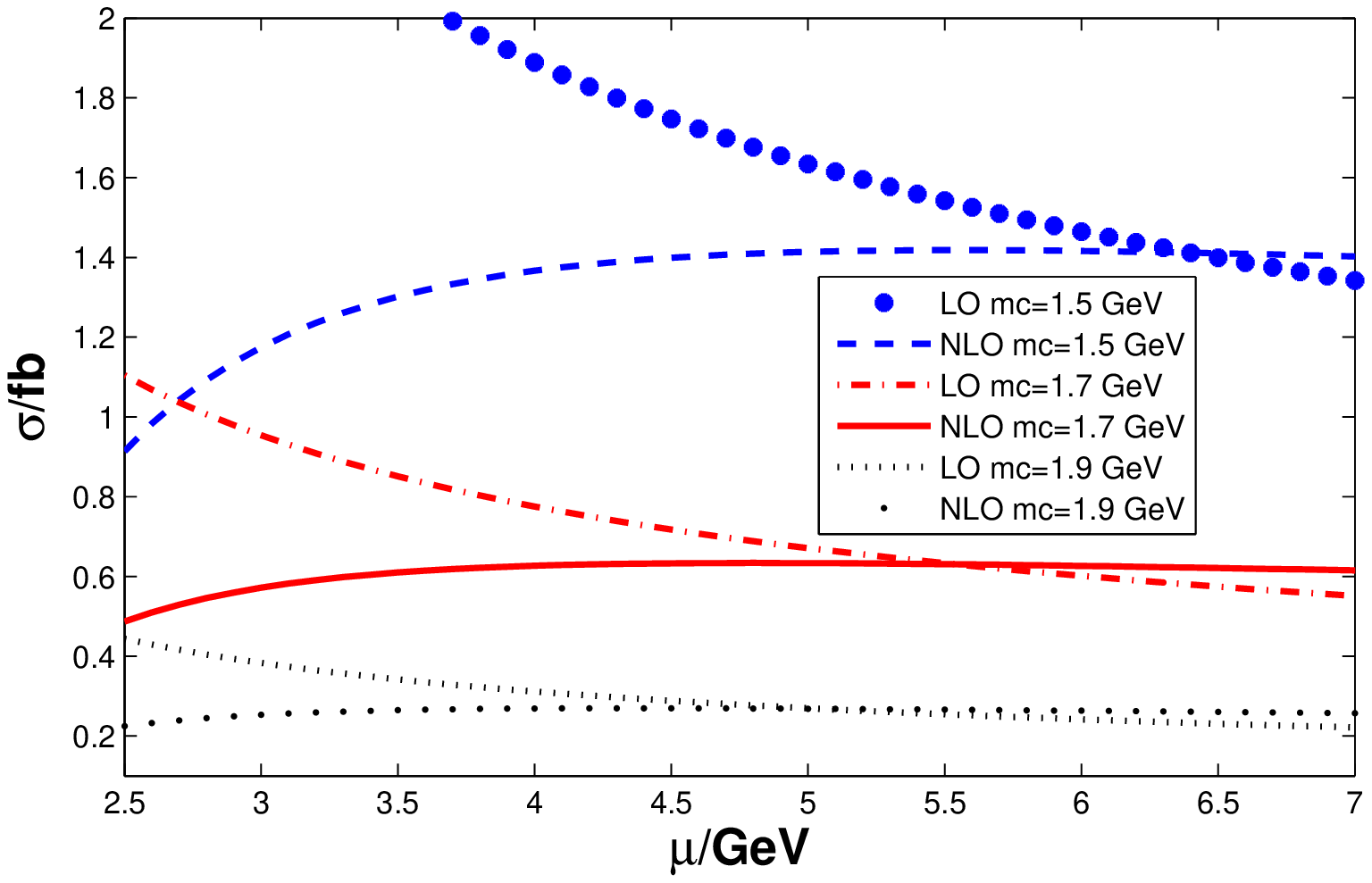}
\figcaption{\label{fig4} LO and NLO cross sections with different $m_c$ versus running scale $\mu$ for $e^+e^-\rightarrow\eta_c+\psi_{2}(1D)$.}
\end{center}

In Figs. \ref{fig5}-\ref{fig7}, we present the LO and NLO cross sections for $e^+e^-\rightarrow\eta_c+h_c(1P)/\psi_{1,2}(1D)$ versus center-of-mass energy ($E_{cm}$=$\sqrt{s}$), respectively, where scale $\mu$ and mass $m_c$ are fixed. The figures show that NLO correction for $\eta_c+h_c(1P)$ production is positive in the displayed region, while that for $\eta_c+\psi_1(1D)$ is negative. Interestingly, for the $\eta_c+\psi_2(1D)$ channel, the NLO correction has an inversion at $E_{cm}=10.36$ GeV. Going across this energy point, the NLO correction changes from positive to negative. Generally, all curves go down with the increase of center-of-mass energy. It is worth noting that NRQCD factorization formalism for double charmonium only holds when  $\sqrt{s} \gg m_c$ \cite{Bodwin:2008nf}, and it may break down at the double charmonium threshold because of the color transfer effect \cite{Nayak:2007mb}, whereas in drawing Figs. \ref{fig5}-\ref{fig7} the center-of-mass energy has been extended to the threshold region, lower than the typical B-factory energy, just for a schematic display.
\begin{center}
\includegraphics[width=8.5cm]{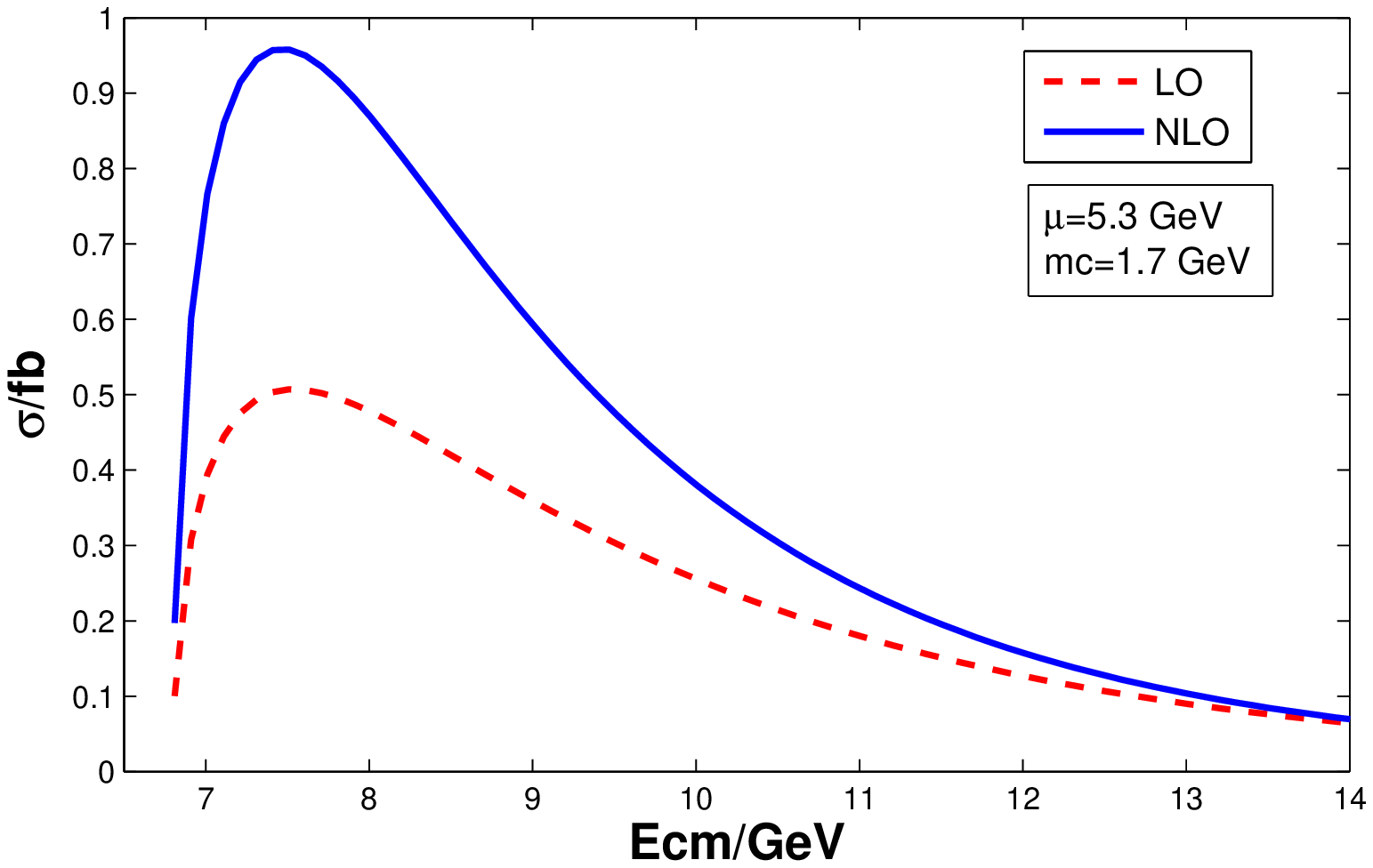}
\figcaption{\label{fig5} LO and NLO cross sections versus running energy of center-of-mass($E_{cm}$) for $e^+e^-\rightarrow\eta_c+h_c(1P)$ with $m_c=1.7$ GeV at $\mu=5.3$ GeV.}
\end{center}
\begin{center}
\includegraphics[width=8.5cm]{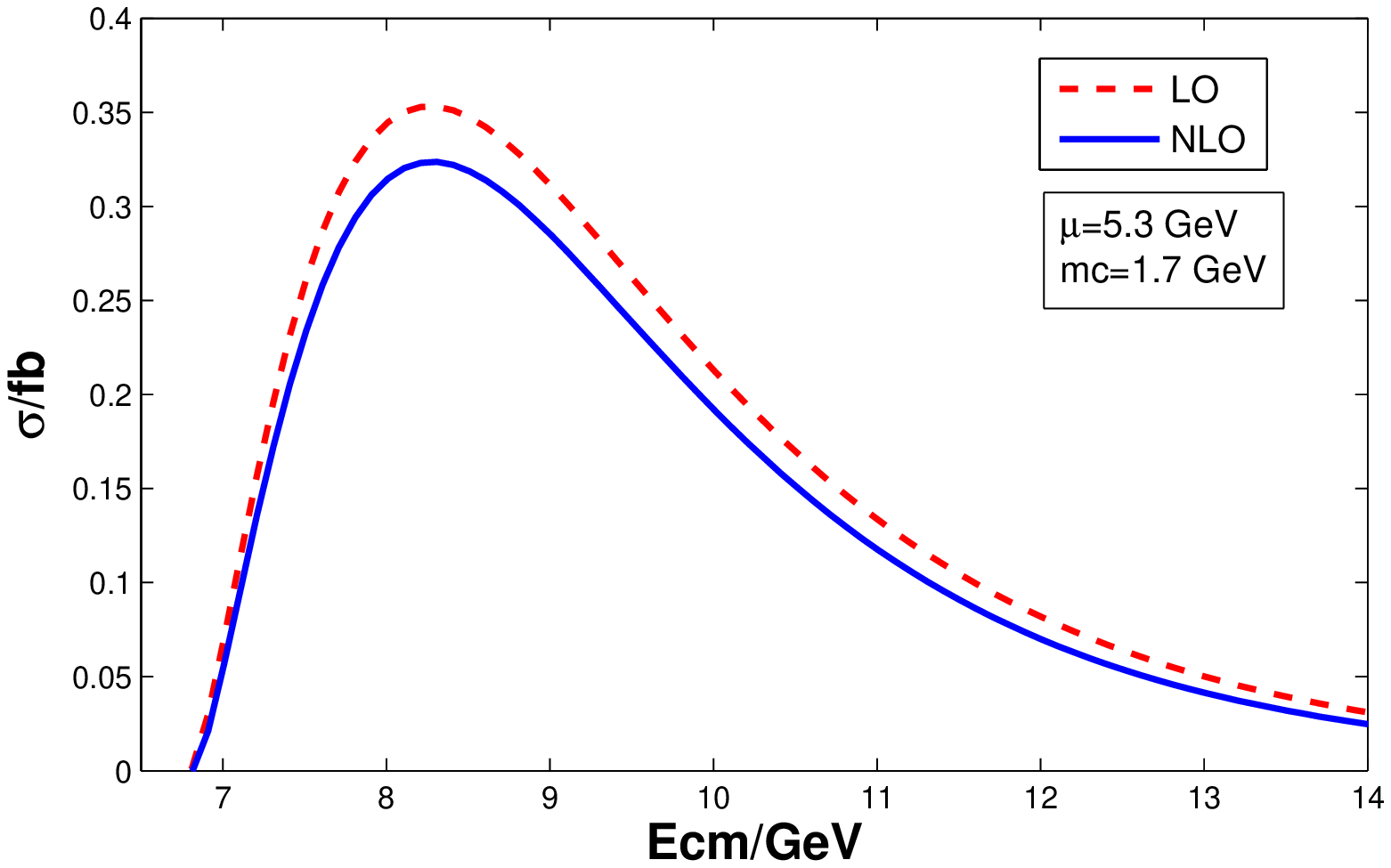}
\figcaption{\label{fig6} LO and NLO cross sections versus running energy of center-of-mass($E_{cm}$) for $e^+e^-\rightarrow\eta_c+\psi_1(1D)$ with $m_c=1.7$ GeV at $\mu=5.3$ GeV.}
\end{center}
\begin{center}
\includegraphics[width=8.5cm]{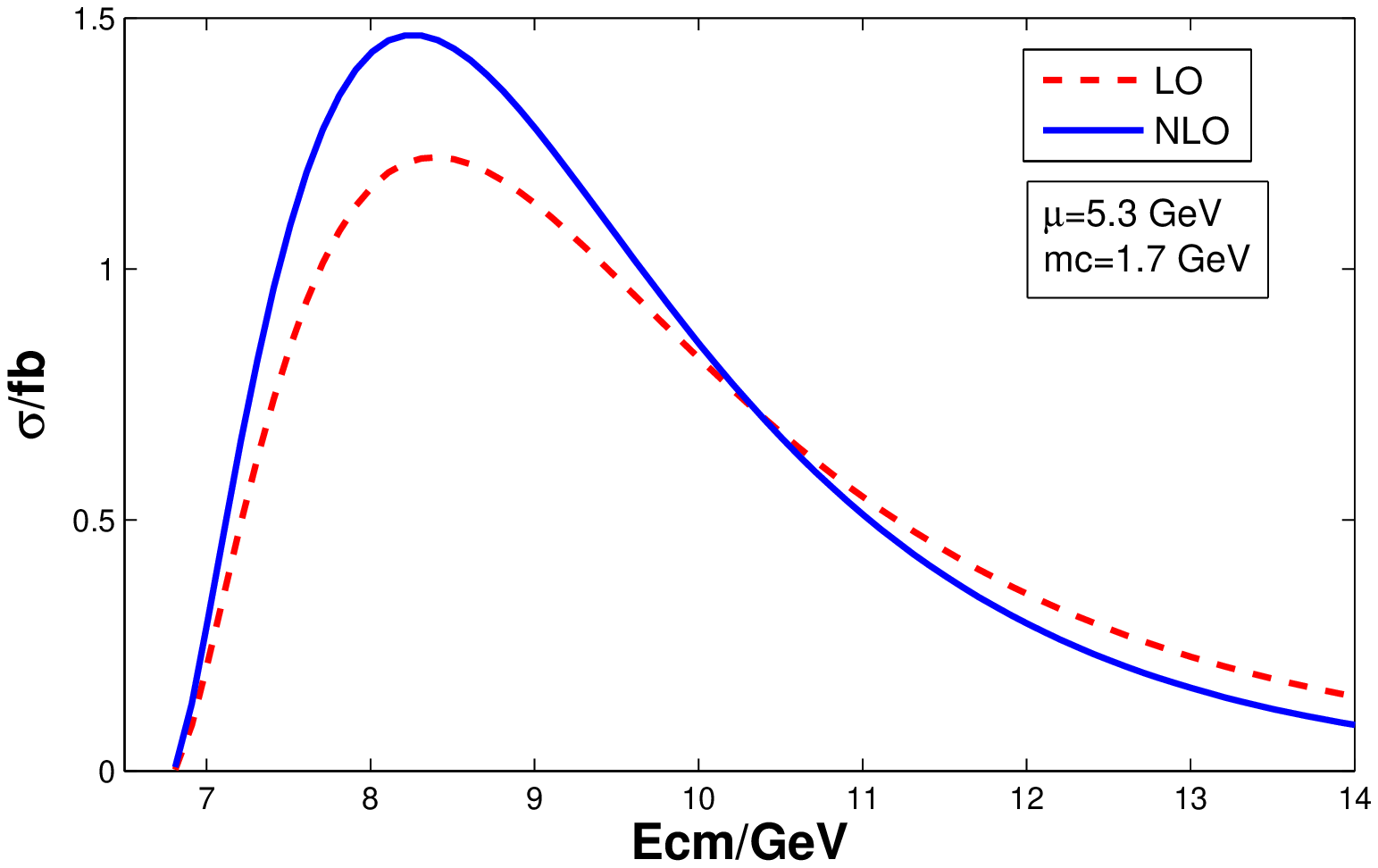}
\figcaption{\label{fig7} LO and NLO cross sections versus running energy of center-of-mass($E_{cm}$) for $e^+e^-\rightarrow\eta_c+\psi_2(1D)$ with $m_c=1.7$ GeV at $\mu=5.3$ GeV.}
\end{center}

Since $M_{\psi_{1,2}}> M_{D\bar{D}}$, i.e. it is above the open charm threshold, and $\eta_c+\psi_{1,2}(1D)$ cross sections are at the order of $1$ fb, in experiment it turns out to be harder to detect $\psi_{1,2}$ than $h_c$ through double charmonium processes at super B-factories. For $\eta_c+h_c(1P)$ production, defining the relative production ratio R as
\begin{eqnarray}
R=\frac{\sigma[e^+e^-\rightarrow\eta_c+h_c]} {\sigma[e^+e^-\rightarrow\eta_c+J/\psi]},
\end{eqnarray}
and considering the Belle data of $\sigma[e^+e^-\rightarrow\eta_c+J/\psi] = 33$ fb \cite{Abe:2002rb}, we have $R=0.010(0.009)$ at $\mu=2 m_c(\sqrt{s}/2)$. If the BaBar data  $\sigma[e^+e^-\rightarrow\eta_c+J/\psi]=17.6$ fb \cite{Aubert:2005tj} are adopted, we then have $R=0.019(0.017)$ at $\mu = 2m_c(\sqrt{s}/2)$. If we adopt the NLO results of Ref. \cite{Zhang:2005cha} for $\sigma[e^+e^-\rightarrow\eta_c+J/\psi]=18.9$ fb, then $R=0.018$ under the same renormalization scale $\mu=2m_c$. In future super B-factories, this ratio $R$ may stand as a benchmark for the estimation of the possibility of observing the $h_c(1P)$ through double charmonium process.

\section{Summary and Conclusions}

In this work, we have studied the double charmonium processes $e^+e^-\rightarrow\eta_c+h_c(1P)/\psi_{1,2}(1D)$ at NLO accuracy  under the NRQCD factorization mechanism. Cross sections with varying  charm quark mass $m_c=1.7\pm0.2$ GeV at typical renormalization scales ($\mu=2m_c, \sqrt{s}/2$) were analyzed in detail. The magnitudes of cross section versus energy scale $\mu$ and center-of-mass energy $\sqrt{s}$ at LO and NLO were evaluated. We have also estimated the relative production rate of $R=\sigma[e^+e^-\rightarrow\eta_c+h_c]/\sigma[e^+e^-\rightarrow\eta_c+J/\psi]$, which might be helpful for the measurement of double charmonium exclusive production in future super B-factories.

Through our study, we find that the NLO corrections for $e^+e^-\rightarrow\eta_c+h_c(1P)/\psi_{1,2}(1D)$  are small, and the convergence of perturbative expansion works well up to NLO. When the scale $\mu$ lies in the region $[2m_c, \sqrt{s}/2]$, the NLO correction for $\eta_c+h_c(1P)$ production is positive, yet it is negative for $\eta_c+\psi_{1,2}(1D)$. The scale dependence has been clearly reduced when NLO corrections are taken into account. At $\sqrt{s}=10.6$ GeV, the relation $\sigma_{\psi_1}<\sigma_{h_c}<\sigma_{\psi_2}$ exists at both LO and NLO, no matter whether $\mu=2m_c$ or $\mu=\sqrt{s}/2$. It is worth noting that the cross sections of the  three processes considered are sensitive to the charm quark mass, which hence is the main source of uncertainty in our results.

Although $h_c(1P)$ has been measured through the $h_c \to J/\psi+\pi^0 \to (e^+e^-)+\pi^0$ process, its branching fraction is not determined, and there has been no signal observed in the other $h_c$ primary decay mode $h_c \to J/\psi+2\pi \to (e^+e^-)+2\pi$ \cite{Armstrong:1992ae}. The results in this work may be helpful to $h_c$ and NRQCD studies in B-factories in future.

\end{multicols}

\vspace{2mm}

\centerline{\rule{80mm}{0.1pt}}
\vspace{2mm}

\begin{multicols}{2}

\end{multicols}

\clearpage

\end{document}